\documentclass[12pt,english]{article}
\usepackage[LGR,T1]{fontenc}
\usepackage[latin9]{inputenc}
\usepackage{geometry}
\usepackage{units}
\usepackage{pmboxdraw}
\usepackage{amsmath}
\usepackage{esint}

\makeatletter

\DeclareRobustCommand{\greektext}{%
  \fontencoding{LGR}\selectfont\def\encodingdefault{LGR}}
\DeclareRobustCommand{\textgreek}[1]{\leavevmode{\greektext #1}}
\DeclareFontEncoding{LGR}{}{}
\DeclareTextSymbol{\~}{LGR}{126}
\newcommand{\lyxmathsym}[1]{\ifmmode\begingroup\def\b@ld{bold}
  \text{\ifx\math@version\b@ld\bfseries\fi#1}\endgroup\else#1\fi}

\usepackage{babel}
\date{ }

\makeatother

\usepackage{babel}
\begin{document}

\title{On the material origin of the cosmological constant}

\author{{\normalsize{Herman Telkamp}}%
\thanks{{\footnotesize{Jan van Beverwijckstr. 104, 5017 JA Tilburg, Netherlands,
email: herman\_telkamp@hotmail.com}}%
}}
\maketitle
\begin{abstract}
\noindent We consider a possible connection between matter and cosmological
constant $\Lambda$ via the Newtonian cosmic potential of the matter
within the expanding particle horizon. Consistent with GR, an increasing
potential may drive the metric expansion of space. Cosmic recession
of mass must, in turn, affect the potential in an opposite sense.
Independent of this, several considerations point at{\normalsize{
}}$-\tfrac{1}{2}c^{2}$ as the representation of the background potential
in the various GR metrics. This suggests that the cosmic potential,
while subject to the expansion of space, always yields a constant
background potential $-\tfrac{1}{2}c^{2}$. Analysis of this ``redshift''
of the cosmic potential yields for perfect fluids the exact same solutions
of the scale factor as the standard Friedmann equations, including
an accelerating de Sitter universe. Densities of dust and $\Lambda$
relate as $\Omega_{{\scriptscriptstyle \Lambda}o}=\tfrac{3}{2}\Omega_{mo}$.
Though counter intuitive at first sight, gravity may drive cosmic
acceleration.\bigskip{}

\end{abstract}
The very nature of the cosmological constant $\Lambda$ has been subject
of discussion from its conception onward. Einstein, while convinced
of the Machian principle, held for a long time that the constant must
have a material origin. De Sitter, on the other hand, showed that
his solution of the Einstein equations for an empty universe allows
for a non-zero cosmological constant. That is, for an accelerating
empty universe with $\Lambda>0$, hence attributing the origin of
the constant to space itself. In empty space there is not much one
can relate the energy density of $\Lambda$ to, other than the zero-point
energy of the quantum vacuum. But any attempts to connect this huge
vacuum energy density to the extremely small dark energy density failed
so far. 

Apart from this particular question about the origin of $\Lambda$,
there is no satisfactory answer to why the universe expands anyway.
Recession is viewed either as inertial motion or as metric expansion.
Both views are viable and exchangeable, at least to a considerable
extent, but only if the universe is flat and decelerating. The Newtonian
view on recession is that galaxies move apart, simply because they
did so in the past, as a result of the big bang and/or the inflationary
era. There is no Newtonian answer to an accelerating universe, though.
Most cosmologists today view recession of galaxies as expansion of
space itself. But no physical explanation exists for why this accelerates,
other than by dark energy as a ``repulsive force''. 

Along the line of Einstein, we explore the possibility of a material
origin of the positive cosmological constant. This may seem counter
intuitive, as matter attracts other matter and so seems to oppose
expansion. Yet, matter can cause metric expansion. That is, if the
total gravitating mass increases. This is conceivable if one considers
e.g. the Schwarzschild space outside of an increasing central mass
$M(t)$. The proper displacement $dr$ corresponding to a fixed coordinate
displacement $dR$ at coordinate distance $R$ from $M$ will evolve
according to $dr(t)=dR/\left(1-2GM(t)/c^{2}R\right)^{-\unitfrac{1}{2}}$,
so $dr$ will increase along with $M$. And actually, something similar
is taking place in the universe as the total mass inside of the expanding
particle horizon continually increases \cite{Margalef,Davis}.%
\footnote{Total mass within the particle horizon increases, since the comoving
distance to this horizon always increases, up to a possible asymptotic
convergence. See e.g. \cite{Margalef,Davis}.%
} Thus expansion of the particle horizon provides a premise for metric
expansion by gravity. 

To consider this further, we assume the spacetime metric of the universe
is ruled by the cosmic gravitational potential. But, how to actually
determine the cosmic potential is a relatively seldom addressed and
largely unanswered question. There exists a simple indication however.
Consider the Schwarzschild metric with the sun as central mass. Note
that the Schwarzschild coefficient $\alpha_{s}=1-2GM_{sun}/c^{2}R$
is virtually unity anywhere outside of the sun (as $2GM_{sun}/c^{2}R<4.3\cdot10^{-6},\; R\geq R_{sun})$.
In other words, the sun's potential $\varphi_{sun}=-GM_{sun}/R$ is
negligible, even at its surface, relative to this huge other potential
$\varphi_{o}=-\tfrac{1}{2}c^{2}$, which we can associate with little
else but the background potential of the universe. We will address
this further below.

Ernst Mach pointed out that, while the influence of a remote star
on the total potential here drops off as $1/r$, the number of stars
at distance $r$ is proportional to $r^{2}$. In this view, which
ignores redshift, the farthest stars make (together) the biggest contribution
to the total potential here. This concerns precisely the masses that
just appear within the moving cosmic horizon and add to the total
gravitating mass. Then, will these extra masses also add to the potential?
Not necessarily, since any increase of the potential will increase
metric distances (i.e. cause apparent recession of all cosmic masses),
which in turn reduces the value of the potential. So the effect of
mass increase on potential is suppressed by metric expansion of space.
Then which equations exactly govern this mechanism? We approach this
subject from two directions; a ``Newtonian'' and a ``Machian''
approach, as follows. We will throughout assume a flat universe. Angular
coordinates are omitted, as we are only concerned with radial motion
of cosmic expansion. Subscript $o$ denotes present time (with the
exception of $\varphi_{o}$, which is a constant).

\medskip{}

\noindent \textit{Newtonian cosmic potential} - We assume gravity
propagates at the constant speed of light $c$. Then, in a big bang
scenario, from some initial cosmic time $t=0$ onward, the total Newtonian
potential $\varphi_{{\scriptscriptstyle N}}$ is generated by the
increasing total mass within the steadily expanding particle horizon.
Hence, we assume the gravitational horizon coincides with the particle
horizon%
\footnote{%
\begin{minipage}[t]{1\columnwidth}%
This is arguable, but may be left undecided, since the precise distance
to the gravitational horizon is not relevant here.%
\end{minipage}%
}. Along with the Newtonian potential, scale factor $a$ and proper
distance $dr$ increase, as $dr=ad\chi$, where $\chi$ is the comoving
coordinate. To circumvent the mutual dependency of cosmic potential
and spatial metric, we consider the cosmic potential in a frame of
which the spatial coordinates are independent of the scale factor.
The comoving coordinate $\chi$ fits this purpose, since masses (at
large) are at rest in this coordinate, so their comoving position
is independent of the changing cosmic potential. The particle horizon
$\chi_{ph}$ propagates as 
\begin{equation}
\chi_{ph}=c\eta,\label{eq:Xph-1}
\end{equation}
where $\eta$ is conformal time. As usual, the scale factor is normalized
at $a_{o}=1$ at present time. This is convenient, since the constant
mass density in comoving coordinates $\rho_{\chi}$ (the ``comoving''
density) then equals $\rho_{o}$, the present mass density (ordinary
+ dark matter) in proper units,
\begin{equation}
\rho_{\chi}=\rho_{o}.\label{eq:rhoX}
\end{equation}
The increment of the total mass $M$ enclosed by the moving horizon
is 
\begin{equation}
dM=4\pi\chi^{2}\rho_{\chi}d\chi.\label{eq:dM}
\end{equation}
Then, the Newtonian cosmic potential in the ($\eta,\chi$) frame follows
from the integral 
\begin{equation}
\varphi_{{\scriptscriptstyle N}}(\eta)=-\underset{{\scriptstyle o}}{\overset{{\scriptstyle \chi_{ph}=c\eta}}{{\textstyle \int}}}\:\frac{G}{\chi}\frac{dM}{d\chi}d\chi=-\underset{{\scriptstyle o}}{\overset{{\scriptstyle \chi_{ph}=c\eta}}{{\textstyle \int}}}\:\frac{G4\pi\chi^{2}\rho_{\chi}}{\chi}d\chi=-2\pi G\rho_{\chi}c^{2}\eta^{2}=-2\pi G\rho_{o}c^{2}\eta^{2}.\label{eq:Newton cosmic potential}
\end{equation}

\medskip{}

\noindent \textit{Machian cosmic potential} - In a well known paper
of 1953 \cite{Sciama}, Sciama derived, on theoretical grounds and
several assumptions, a cosmic potential $\sim-c^{2}$. Yet, in a relatively
unknown paper of 1925 \cite{Schroedinger}, Schrödinger presented
a Machian calculation of the anomalous perihelion precession of planetary
orbits, yielding exactly Einstein's famous GR-based result for Mercury.
Assume $m$ is the mass of the planet, orbiting the sun at a separation
$r$. The essence of this calculation is that the planet is assigned
a small additional inertia $\mu$ in the direction of the sun, proportional
to the potential of the sun $\varphi_{sun}(r)$ at the position of
the planet, while the (fixed) Newtonian inertia $m$ of the planet
is considered to be proportional to the (fixed) background potential
$\varphi_{o}$ of the universe, which is represented by the hollow
sphere model. The (very small) Machian inertia $\mu(r)=m\varphi_{sun}(r)/\varphi_{o}$
reproduces the GR expression of the anomalous precession exactly,
provided 
\begin{equation}
\varphi_{o}=-\tfrac{1}{2}c^{2}.\label{eq:phio}
\end{equation}
This is no surprise, since $\varphi_{o}$ equals the potential appearing
in the Schwarzschild metric, as noted above. Yet, the Machian model
unambiguously identifies $\varphi_{o}$ as the cosmic background potential.
Contrary to Sciama's model, Schrödinger's calculation regards perfectly
observable local mechanics of the solar system. The result matches
actual measurements of the perihelion precession of Mercury and other
planets. Therefore, (\ref{eq:phio}) is indirectly confirmed by observation. 

\medskip{}

Thus, the cosmic potential appears as a constant in our local frame.
From this we infer that metric expansion of the universe turns the
growing Newtonian potential $\varphi_{{\scriptscriptstyle N}}(\eta)$
into a constant background potential $\varphi'_{{\scriptscriptstyle N}}(\eta)=\varphi_{o}$
in our proper frame. The obvious analogy is the speed of light in
the gravitational field. It is well known that, to a remote observer,
light appears to slow down while traveling through a region of stronger
potential (i.e. Shapiro delay). However, the observer near the gravitating
object will always measure exactly $c$, locally. This means that
the proper coordinates are being adjusted by the local potential in
such a way that local measurement of the speed of light always yields
$c$. Likewise, the local observer will always experience a fixed
background potential $\varphi_{o}$. 

The background potential $\varphi_{o}$ also offers a natural interpretation
of the coefficient $\alpha_{s}$ in the Schwarzschild metric,
\begin{equation}
ds^{2}=\alpha_{s}c^{2}dT^{2}-\frac{dR^{2}}{\alpha_{s}}-R^{2}d\Omega^{2}.\label{eq:Schwarzschild metric-1}
\end{equation}
Using $\varphi_{o}=-\tfrac{1}{2}c^{2}$ and $\varphi_{{\scriptscriptstyle M}}(R)=-GM/R$,
and assuming ${\scriptstyle \text{\textSFxi}}\varphi_{{\scriptscriptstyle M}}(R){\scriptstyle \text{\textSFxi}}\ll\text{\ensuremath{{\scriptstyle \lyxmathsym{\textSFxi}}}}\varphi_{o}{\scriptstyle \text{\textSFxi}}$
the coefficient $\alpha_{s}$ can be expressed as 
\begin{equation}
\alpha_{s}=1-\frac{2GM}{c^{2}R}=\frac{\varphi_{o}-\varphi_{{\scriptscriptstyle M}}(R)}{\varphi_{o}}\simeq\frac{\varphi_{o}}{\varphi_{o}+\varphi_{{\scriptscriptstyle M}}(R)}=\frac{\varphi(\infty)}{\varphi(R)},\label{eq:alpha interpretation}
\end{equation}
i.e. as the ratio of the total potential $\varphi(\infty)=\varphi_{o}$
at infinity and the total potential $\varphi(R)=\varphi_{o}+\varphi_{{\scriptscriptstyle M}}(R)$
at $R$. This makes perfectly sense from a Machian point of view.
A non-zero value of the potential at infinity may raise some eyebrows
though, as the Schwarzschild space is asymptotically Minkowski space,
which is supposed to be empty. Yet, considering (\ref{eq:alpha interpretation}),
the apparent background potential in the Schwarzschild space suggests
that Minkowski space is actually representing our flat cosmological
background. That is, a universe homogeneously filled with matter,
producing a flat potential $\varphi_{o}=-\tfrac{1}{2}c^{2}$. This
potential is represented in the metric coefficients of the various
GR metrics (or hidden as $\alpha_{s}=1$ in the Minkowski metric,
being the limit of the Schwarzschild metric, i.e. ${\scriptstyle \underset{{\scriptscriptstyle M}\rightarrow0}{\lim}}(\varphi_{o}-\varphi_{{\scriptscriptstyle M}}(R))/\varphi_{o}=1$.
The constant background potential $\varphi_{o}$ provides a material
backplane to both ``empty'' Minkowski space and the Schwarzschild
``vacuum''. Note that there is nothing in GR that contradicts or
excludes this interpretation. A homogeneous, isotropic matter distribution
does nothing but creating a flat background, like Minkowski space.
It is not a coincidence that the FRW metric of our universe is conformally
flat, i.e. Minkowskian for any particular value of the scale factor.
Actually, a non-empty Minkowski space solves a few long lasting problems:
a flat material background reconciles GR with the Machian principle,
as Schrödinger showed in his calculation of the anomalous precession.
Moreover, it releases Minkowski space from the unfavorable connotation
of being an absolute space. With the cosmic background in place, any
motion in Minkowski space becomes relative, as is most desirable from
a relativistic perspective. This has some notable consequences. Of
particular interest here: if Minkowski space is non-empty, then this
suggests that the de Sitter universe may not be empty too. We will
substantiate this by showing that, without invoking exotic dark energy,
a de Sitter space emerges naturally from a universe that contains
nothing but matter (``dust''). The metric evolution is hidden in
the ``redshift'' of the growing Newtonian potential to the constant
Machian potential. 

\medskip{}

\noindent \textsl{The Newtonian potential subject to metric expansion}
- In light of the above, we require this ``redshifted'' value $\varphi'_{{\scriptscriptstyle N}}\equiv\varphi_{{\scriptscriptstyle N}}/f(a)$
to come out equal to the Machian potential, i.e. $\varphi'_{{\scriptscriptstyle N}}=-\tfrac{1}{2}c^{2}$,
or 
\begin{equation}
\varphi'_{{\scriptscriptstyle N}}(\eta)=\frac{\varphi_{{\scriptscriptstyle N}}(\eta)}{f(a)}=-2\pi G\rho_{\chi}c^{2}\frac{\eta^{2}}{f(a)}=-\tfrac{1}{2}c^{2}.\label{eq:phi'}
\end{equation}
Normalizing $f(a_{o})=1$ gives 
\begin{equation}
\eta_{o}^{2}=\frac{1}{4\pi G\rho_{\chi}},\label{eq:nuo}
\end{equation}
and
\begin{equation}
f(a)=\frac{\eta^{2}}{\eta_{o}^{2}}.\label{eq:f(a)}
\end{equation}
Like for light, one expects the redshift factor to be a simple function
$f(a)$ of the scale factor, which is an immediate measure of the
state of metric expansion. Consider the Friedmann equation of the
\textgreek{L}CDM model
\begin{equation}
\frac{\dot{a}^{2}}{a^{2}}=H^{2}=H_{o}^{2}(\frac{\Omega_{mo}}{a^{3}}+\Omega_{{\scriptscriptstyle \Lambda}o}),\label{eq:LCDM Friedmann}
\end{equation}
where $\Omega_{mo}$, $\Omega_{{\scriptscriptstyle \Lambda}o}$ and
$H_{o}$ are present values of matter density, dark energy density
and the Hubble parameter, respectively. For simplicity, we ignore
other density terms. The model (\ref{eq:LCDM Friedmann}) describes
a universe filled with a mixture of two perfect fluids: dust and dark
energy (in the form of the cosmological constant), with equation of
state parameters $w=0$ and $w=-1$, respectively. For a flat universe,
the deceleration parameter $q$ relates to $w$ according to
\begin{equation}
q=\tfrac{1}{2}(1+3w).\label{eq:q=00003Df(w)}
\end{equation}
At early times, matter dominates and $w=0$, $q=1/2$. At late times
$\Lambda$ dominates and $w=-1$, $q=-1$. So $q(t)$ runs from $1/2$
to $-1$. The solution of the Friedmann equation for a perfect fluid
with equation of state parameter $w$ \cite{Gron} is, when expressed
in terms of the deceleration parameter $q$, 
\begin{equation}
a(t)=\begin{cases}
\:{\textstyle \Bigl({\displaystyle \frac{t}{t_{o}}}\Bigr)^{{\scriptstyle 1/(1+q)}}} & \quad q\neq-1,\quad\quad t_{o}={\displaystyle \frac{1}{(1+q)H_{o}}=\frac{\eta_{o}}{1+q}},\\
\\
\;\: e^{{\displaystyle {\scriptstyle H_{{\scriptscriptstyle \Lambda}}t}}} & \quad q=-1,\quad\quad H_{{\scriptscriptstyle \Lambda}}^{2}=\Lambda c^{2}/3=1/\eta_{o}^{2}.
\end{cases}\label{eq:a(t) q}
\end{equation}
This solution allows us to express conformal time as a function of
scale factor, thus to solve for redshift factor $f(a)$, according
to
\begin{equation}
f(a)=\frac{\eta^{2}}{\eta_{o}^{2}}=\biggl(\frac{1}{\eta_{o}}\intop_{t_{o}}^{t}a^{-1}dt+1\biggr)^{2}=\begin{cases}
\;\; a^{2q} & \quad q\neq-1\\
\;(2-a^{-1})^{2} & \quad q=-1.
\end{cases}\label{eq:f(a)-1}
\end{equation}

\noindent So this is the factor which turns the evolving Newtonian
potential in a constant background potential $\varphi_{o}$. Differentiating
the square root of (\ref{eq:f(a)-1}) gives, for any $q$,
\begin{equation}
d\eta=\pm\eta_{o}qa^{q-1}da.\label{eq:dnu-1}
\end{equation}
Substituting $d\eta\equiv dt/a$ in (\ref{eq:dnu-1}) yields the differential
equation for the scale factor
\begin{equation}
\frac{da}{dt}=\pm\frac{a^{-q}}{q\eta_{o}},\label{eq:metric ODE a(t)}
\end{equation}
which has (\ref{eq:a(t) q}) as solutions, as expected. 

\medskip{}

\noindent Hence, in case of a \textit{matter-only} universe, the constancy
of the background potential, subject to expansion of the horizon,
gives rise to eq. (\ref{eq:metric ODE a(t)}), which reproduces the
standard solutions (\ref{eq:a(t) q}) of the Friedmann equation for
a perfect fluid of arbitrary equation of state $w$, including the
solution of a de Sitter universe for $w=-1$ (i.e. $q=-1$). The present
value $\Omega_{{\scriptscriptstyle \Lambda0}}$ of the density parameter
of a material $\Lambda$ follows from (\ref{eq:nuo}),(\ref{eq:rhoX})
and (\ref{eq:a(t) q}),

\begin{equation}
\Omega_{{\scriptscriptstyle \Lambda0}}=\frac{H_{{\scriptscriptstyle \Lambda}}^{2}}{H_{o}^{2}}=\frac{4\pi G\rho_{\chi}}{H_{o}^{2}}=\frac{4\pi G\rho_{o}}{H_{o}^{2}},\label{eq:Ohm Lambda}
\end{equation}
which unambiguously shows its connection with matter. This leads us
to conclude that the cosmological constant may indeed have a material
origin, as conjectured by Einstein. Furthermore, since the density
parameter of dust has a present value
\begin{equation}
\Omega_{mo}=\frac{\rho_{o}}{\rho_{co}}=\frac{8\pi G\rho_{o}}{3H_{o}^{2}},\label{eq:ohm m}
\end{equation}
it follows that the two density parameters have a fixed ratio,%
\footnote{This fixed ratio holds for any value of present time $t_{o}$. For
a particular $t_{o}$, comoving density $\rho_{\chi}$ is constant
over time. But it scales like $\rho_{o}$ with change of present time
$t_{o}$, since $\rho_{\chi}=\rho_{o}$ for normalized scale factor
$a(t_{o})=a_{o}=1$. %
} 
\begin{equation}
\Omega_{{\scriptscriptstyle \Lambda}o}=\tfrac{3}{2}\Omega_{mo}.\label{eq:Ohm m / ohm lambda}
\end{equation}
For $\Omega_{r}=\Omega_{k}=0$ (no radiation, flat universe) this
gives exactly $\Omega_{mo}=0.4$ and $\Omega_{{\scriptscriptstyle \Lambda}o}=0.6$.
Interestingly close to observation, though not quite good enough,
given the current status of observed values ($\widetilde{\Omega}_{mo}\simeq0.3$,
$\widetilde{\Omega}_{{\scriptscriptstyle \Lambda}o}\simeq0.7$). So
this presents an issue, with no obvious resolution. 

One however may speculatively consider that a material origin of $\Lambda$
could impact the structure of the \textgreek{L}CDM model (\ref{eq:LCDM Friedmann}).
The density parameters in the standard Friedmann equation are assumed
to represent distinct forms of energy and to be independent, which
is not the case with the proposed connection of $\Lambda$ to matter.
Dependency gives rise to cross-terms, hence, to a modification of
the standard model by an additional cross term, that could possibly
make up for the (small) mismatch of the density parameters.

\medskip{}

The de Sitter universe is an asymptotic solution of the Friedmann
equations with positive cosmological constant. Asymptotically $a(t)$
approaches infinity and matter density obviously goes to zero. So,
is the de Sitter space eventually empty nevertheless? It may appear
so, but if the cosmological constant is indeed connected to the cosmic
potential, then it needs matter to exist at all.

\end{document}